# COSMOLOGICAL IMPLICATIONS OF THE NEW METRIC FOR AN ACCELERATING EXPANDING, AND DOUBLY ROTATING UNIVERSE

**Evangelos Chaliasos**

365 Thebes Street

GR-12241 Aegaleo

Athens, Greece

***Abstract***

We give the correct interpretation of the new metric, found lately by the author. This metric results as an exact solution of the Einstein field equations, without the cosmological constant. The new feature is the introduction, from the beginning, of two rotations attributed to the Universe. As a result also an expanding, and indeed in accelerating rate, Universe is obtained. Thus “dark energy” is not necessary at all. We explore some properties of this Universe in the present paper. In particular, all old properties of the Universe concerning the Hubble law remain in a modified way, implying anisotropy. Besides, we explore the validity of the Ryle-Clarke effect, which is verified, and we find the correct age of the Universe. Finally, the resulting shape of the Universe is examined, and it is found to be hypertoroidal.

## 1. Introduction

In our previous paper [1] we considered our three-dimensional space embedded in a fictitious four-dimensional space, which we allowed to perform two independent rotations, each on a coordinate plane, these two planes having in common only the origin. Then we wrote down the appropriate for this case form of the metric in the four-dimensional space-time. After a lot of intermediate calculations, we wrote down the resulting Einstein field equations, which we solved analytically after a lot of effort too.

We thus found that the metric, in comoving coordinates, is

$$ds^2 = c^2dt^2 - U^2(t)\{dr^2 + (d\varphi_1 + \omega_1 dt)^2 + r^2(d\varphi_2 + \omega_2 dt)^2\}, \tag{1.1}$$

describing a differential rotation of the space with angular velocity

$$\omega_1(t) = \frac{c}{E^3}\xi e^{-3Gt}, \tag{1.2}$$

plus a rigid rotation of the space with angular velocity

$$\omega_2(t) = \frac{c}{E^3}\eta e^{-3Gt}, \tag{1.3}$$

in an expanding Universe with scale factor

$$U(t) = \Theta \mathrm{E} e^{Gt}, \tag{1.4}$$

where Θ, E, ξ, η and G are constants coming from integrations. We see at once from (1.4) that the expansion is *accelerated.*

The coordinate system in which the above solution of Einstein´s field equations is described is not "inertial". We can though take from it an "inertial" coordinate system performing the coordinate transformation

$$\begin{Bmatrix} t \\ r \\ \varphi_1 \\ \varphi_2 \end{Bmatrix} \rightarrow \begin{Bmatrix} t \\ r \\ \phi = \varphi_1 + \dfrac{c\xi}{3GE^3} e^{-3Gt} \\ \Phi = \varphi_2 + \dfrac{c\eta}{3GE^3} e^{-3Gt} \end{Bmatrix}. \tag{1.5}$$

Described in the new coordinate system, the metric becomes

$$ds^2 = c^2 dt^2 - U^2(t)\left\{ dr^2 + d\phi^2 + r^2 d\Phi^2 \right\}. \tag{1.6}$$

This metric resembles the flat Friedmann model (space *flat* – spacetime *curved*), but of course this is not that model (look at eq. 1.4). Note that the "inertial" coordinate system is *synchronous*, and thus the proper time τ = t. This is not true for the comoving coordinate system, as we will see in section 2.

In the comoving coordinate system we have for a material point of course

$$\frac{d\varphi_1}{dt} = \frac{d\varphi_2}{dt} = 0. \tag{1.7}$$

Thus from the metric (1.6) we get that in the "inertial" coordinate system the material point performs two rotations, with angular velocities

$$\frac{d\phi}{dt} = \omega_1 \quad \& \quad \frac{d\Phi}{dt} = \omega_2 \quad , \tag{1.8}$$

the first being a *differential* rotation and the second being a *rigid* rotation, or, as we may say , the particle (or the *Universe,* since it consists of all particles) is *doubly* rotating.

In ref. [1], we have found also that, for the energy density, ε = constant×G. That is we had obtained a *steady state* for the Universe. Then, we had commented as follows: since 1) from the Ryle-Clarke effect a steady state has been excluded (for ε ≠ 0 of course), and 2) from the time-symmetric theory of the author (for a brief account of it see e-Appendix B in ref. [1]) we have to have ε = 0, we were obliged to take G = 0, with the result to obtain in addition a *constant* gravitational field, that is *no expansion* at all, and besides, because of the form of the metric (1.6), merely a flat (Minkowskian) space-time (we had attributed the galactic recession, and especially in accelerating rate, to other reasons, as it can be seen in ref. [1] after p. 55).

In ref. [1] the coordinate r was named z. And since it was a comoving coordinate, a material particle had to move accelerating, in the “inertial” coordinate system, on a z = constant plane. Its velocity was proportional to its distance (there) r from the z-axis, and it had to go to a maximum distance r = z, where its velocity became equal to c. It could of course start from r = 0, that is from the z-axis. And since it was valid for all z, the big-bang had to be the whole z-axis. But the big-bang is very well established up to now to be a point rather than an axis. Thus we have to abandon that picture, and accept $G \neq 0$, in order to recover the expansion (from a point of course, the origin).

But then besides the desired expansion (and indeed accelerating), we have to now accept $\varepsilon$ = constant $\neq 0$, that is a steady state. We will see that the author´s time symmetric theory mentioned before eventually is not violated in section 6. Concerning the Ryle-Clarke effect we will try to explore its validity, and we surprisingly see that, because of the new geometry, it is also fulfilled , after a derivation in section 3.

In sections 4 & 5 we will see that Hubble´s law continues to hold, if we neglect the angular velocities $\omega_1$ & $\omega_2$. And, in section 6, we will see that if we take them into account, an anisotropy in the Hubble flow results, even to the first order both in $\omega_1^2$ & $\omega_2^2$, and in the distance from the source, holding thus for small distances and small $\omega_1^2$ & $\omega_2^2$, too.

And a few words on the interpretation of the new metric, in particular in the form (1.6). At first, we see that it describes a *rigid* rotation about the origin on the (r, Φ) plane with angular velocity $\omega_2$. Then the whole (r, Φ) plane rotates *differentially* with angular velocity $\omega_1$ about an axis, say Z, perpendicular to the plane of φ´s, which axis we can take as the common original axis for measuring Φ´s on the (r,Φ) planes (Fig.1)

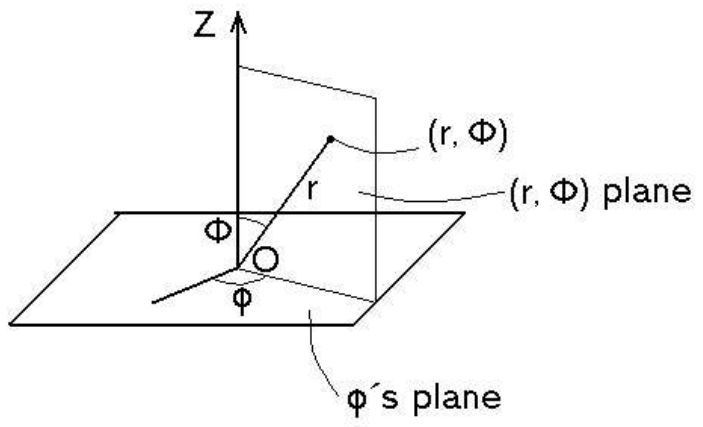

Fig. 1

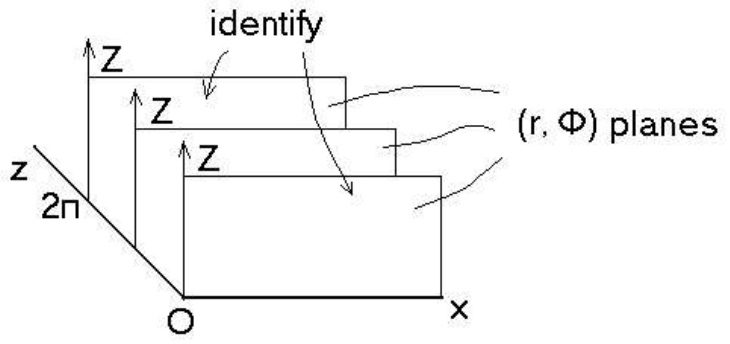

Fig. 2

Alternatively, we can rename (because of the absence of coefficient in $d\varphi^2$) φ to z, so that it is recognized that the metric inside the curly brackets in (1.6) is just the line element, say $d\sigma^2$, of *flat space* in *cylindrical coordinates,* and we may write

$$d\sigma^2 = dr^2 + r^2 d\Phi^2 + dz^2. \tag{1.9}$$

We have to mention however that z (≡ φ) cannot take all real vales, since it represents actually an *angle* (φ), and thus it can take only the values from 0 to 2π, where it must be meant that the value 2π is identified with the value 0. In this picture we may imagine the planes (r, Φ) as being parallel to one another , and perpendicular to the z axis from z = 0 to z = 2π, where the plane for z = 2π is identified with the plane z = 0 (see Fig. 2). Obviously, a particle moves on a solid helix with axis the z-axis, in this 2nd picture.

For more on the correct interpretation of the coordinates, and the resulting geometrical shape of the Universe see Appendix B.

## 2. Proper time

From (1.1) we get

$$g_{00} = 1 - U^2(t)\frac{\omega_1^{\,2}(t)}{c^2} - U^2(t) r^2 \frac{\omega_2^{\,2}(t)}{c^2}, \tag{2.0}$$

or, taking in mind (1.2), (1.3), and (1.4),

$$g_{00} = 1 - \frac{\Theta^2}{E^4}\left(\xi^2 + \eta^2 r^2\right) e^{-4Gt} \tag{2.1}$$

$$\equiv 1 - A^2(r) e^{-4Gt}. \tag{2.2}$$

Thus

$$d\tau = dt\sqrt{1 - A^2(r)e^{-4Gt}} \tag{2.3}$$

so that

$$\tau = \int \sqrt{1 - A^2(r)e^{-4Gt}}\,dt. \tag{2.4}$$

After some calculations, taking also in mind formula 2.284 of [3], we find

$$\tau = -\frac{1}{2G}\sqrt{1 - A^2 e^{-4Gt}} - \frac{1}{2G}\ln\left|\frac{\sqrt{1 - A^2 e^{-4Gt}} - 1}{\sqrt{1 - A^2 e^{-4Gt}} + 1}\right|. \tag{2.5}$$

Obviously the quantity under the square root sign has to be positive, in order for the coordinate system to be realizable by real bodies. Thus, for a certain time t, there is a maximum allowed value of r. At greater r the bodies realizing the coordinate system have to move simply with velocities greater than c, which is absurd. For a fixed r on the other hand, there is a minimum allowed value of t, namely it must

$$t \geq \frac{\ln A(r)}{2G}, \tag{2.6}$$

for which we take from (2.5) that $\tau = 0$. It has no meaning to consider $\tau < 0$, because then the body at r must move again with a velocity greater than c.

## 3. The Ryle-Clarke effect

For simplicity, we assume that $\omega_1 \cong 0 \cong \omega_2$ ,which can be achieved by taking $\xi \cong 0 \cong \eta$. This is justified from the fact that, if $\omega_1$ and $\omega_2$ had appreciable values, then we would have discovered the (double) rotation of the Universe earlier. Thus we will take as the metric in comoving coordinates the metric

$$ds^2 \cong c^2dt^2 - U^2(t)\left\{dr^2 + d\varphi_1{}^2 + r^2 d\varphi_2{}^2\right\}, \tag{3.1}$$

which is the same with the "inertial" metric (1.6). Now since the spatial metric (the metric inside the curly brackets) is flat, we can assume that we have taken the origin at the observer.

The element of spatial (proper) volume will be

$$dV = U^3(t) r dr d\varphi_1 d\varphi_2 \tag{3.2}$$

(see Appendix A, formula (A23) with $\omega_1 = 0 = \omega_2$). Then the number of galaxies dm in this volume will be given by [1]

$$\frac{dm}{dV} = \varepsilon = \frac{6G^2}{kc^2} \quad \text{(dust).} \tag{3.3}$$

Integrating we have

$$m = \varepsilon \int dV, \tag{3.4}$$

or, for the volume of a sphere up to distance l = UR,

$$m = \frac{6G^2}{kc^2} \int_{sphere} U^3(t) r dr d\varphi_1 d\varphi_2 \tag{3.5}$$

$$= \frac{6G^2}{kc^2} (2\pi)^2 \int_0^R U^3(t) r dr \tag{3.6}$$

(see also below, inside the parenthesis after (3.17)). From the metric (3.1) we find that for light (ds = 0) travelling radially ($d\varphi_1 = 0 = d\varphi_2$) towards the origin (the observer)

$$dr = -\frac{cdt}{U(t)}. \tag{3.7}$$

From (3.7) we get

$$r = -\frac{c}{\Theta E} \int e^{-Gt} dt, \tag{3.8}$$

which gives

$$r = \frac{c}{\Theta EG} e^{-Gt} - \frac{c}{\Theta EG} e^{-Gt_0} \tag{3.9}$$

and we obtain

$$U(t) = \frac{c}{G} \left[ r + \frac{c}{G} \frac{1}{U(t_0)} \right]^{-1}. \tag{3.10}$$

Thus, (3.6) gives

$$m = \frac{6c}{kG} (2\pi)^2 \int_0^R \frac{rdr}{(r+\alpha)^3} \quad ; \quad \alpha = \frac{c}{G} \frac{1}{U(t_0)}, \tag{3.11}$$

so that we finally find

$$m = \frac{6c}{kG}(2\pi)^2 \frac{R^2}{2\alpha(R+\alpha)^2}, \tag{3.12}$$

that is we find m as a function of R, say m = f(R).

Now, for a given instant of time (dt = 0), the line element (3.1) gives

$$ds^2 = -U^2(t)\left\{ dr^2 + r^2 d\varphi_2{}^2 + d\varphi_1{}^2 \right\}, \tag{3.13}$$

so that for the element of spatial distance dl we will have

$$dl^2 = U^2(t)\left\{ dr^2 + r^2 d\varphi_2{}^2 + d\varphi_1{}^2 \right\}, \tag{3.14}$$

which can be written as

$$dl^2 = (Udr)^2 + (Urd\varphi_2)^2 + (Ud\varphi_1)^2 \tag{3.15}$$

(Pythagorean theorem). From (3.15) it is evident that the surface element on a sphere centered at the origin and of "radius" r will be given (dr = 0) by

$$dS = (Urd\varphi_2)(Ud\varphi_1) = U^2 r d\varphi_1 d\varphi_2. \tag{3.16}$$

Thus we get for the area of the surface of a sphere centered at the origin (the observer), and of radius r,

$$S_{sphere} = \iint U^2 r d\varphi_1 d\varphi_2 = (2\pi)^2 U^2 r. \tag{3.17}$$

(The volume element then of a spherical shell limited between the above spherical surface and the one of a sphere centered again at the origin and of "radius" r + dr will be (see (3.15))

$$dV = (Udr)S_{sphere} = (2\pi)^2 U^3 r dr, \tag{3.18}$$

with the result that we obtain for the volume of a sphere centered at the origin and of "radius" R

$$V = \int_{sphere} (2\pi)^2 U^3 r dr = (2\pi)^2 \int_0^R U^3(t) r dr \tag{3.19}$$

(cf. (3.5) & (3.6))).

Proceeding in a fashion analogous to §3.6 of [5], we write directly his result for the *flux density* $F(\nu_0)$ of radio signals emitted from a source, of luminosity L, located "distance" R apart with frequency ν, and received by the observer at the frequency $\nu_0$,

$$F(\nu_0) = \frac{LJ(\nu_0 \cdot 1 + z)}{(1+z)\cdot(2\pi)^2 U_0^{\,2} R}, \tag{3.20}$$

where J is the intensity function, and z the redshift, with one *crucial* difference: *we do not write the Euclidean value* $4\pi U_0^2 R^2$ *for the surface of a sphere centered at the source and of radius our "distance" R, but rather the correct for our case value* $(2\pi)^2 U_0^2 R$ (cf. 3.17). Then, since

$$1 + z = \frac{U_0}{U}, \tag{3.21}$$

where (from (3.10)) we have

$$U(t) = \frac{c}{G}\frac{1}{R+\alpha}, \tag{3.22}$$

from (3.20) we get easily the formula we need:

$$F_0 = \frac{LJc}{4\pi^2 G U_0^{\,3} R(R+\alpha)}. \tag{3.23}$$

Thus we found $F_0$ as a function of R, too, say $F_0 = g(R)$.

Thus, we have found the number of radio-galaxies m inside a sphere of "radius" R, i.e. brighter than those with flux density $F_0$, as a function of $F_0$, which function is given parametrically, with parameter the "distance" R, by (3.12) & (3.23), which is of the form

$$\left.\begin{aligned} m &= f(R) \quad &\text{(a)} \\ F_0 &= g(R). \quad &\text{(b)} \end{aligned}\right\} \tag{3.24}$$

This relation has to fit the data of the empirical relation logm = F($\log F_0$), that is to conform with the Ryle-Clarke effect [6, 2].

In fact, from (3.12) we get

$$\log m = \log\left[\frac{6c}{kG}\frac{(2\pi)^2}{2\alpha}\right] + 2\log R - 2\log(R+\alpha), \tag{3.25}$$

while from (3.23) we get

$$-2\log F_0 = -2\log\left(\frac{LJc}{4\pi^2 G U_0^{\,3}}\right) + 2\log R + 2\log(R+\alpha). \tag{3.26}$$

Eliminating R between (3.25) & (3.26),and if we set

$$x \equiv \log m \quad \& \quad y = \log F_0, \tag{3.27}$$

we are finally left with the relation

$$y = x/2 + 2\log\left(10^{-0.5x} - C_1\right) + C_2, \tag{3.28}$$

where $C_1$ & $C_2$ are constants, given by

$$C_1 = \frac{1}{2\pi}\sqrt{\frac{k}{3U_0}} \tag{3.29}$$

and

$$C_2 = \log\left(\frac{1}{2\pi}\frac{G}{c}\sqrt{\frac{3}{kU_0}}LJ\right). \tag{3.30}$$

With some suitable choices of the constants $C_1$ & $C_2$ , we plotted the function (3.28), with the help of *"Mathematica"*, which we expose in Fig. 3. The observational

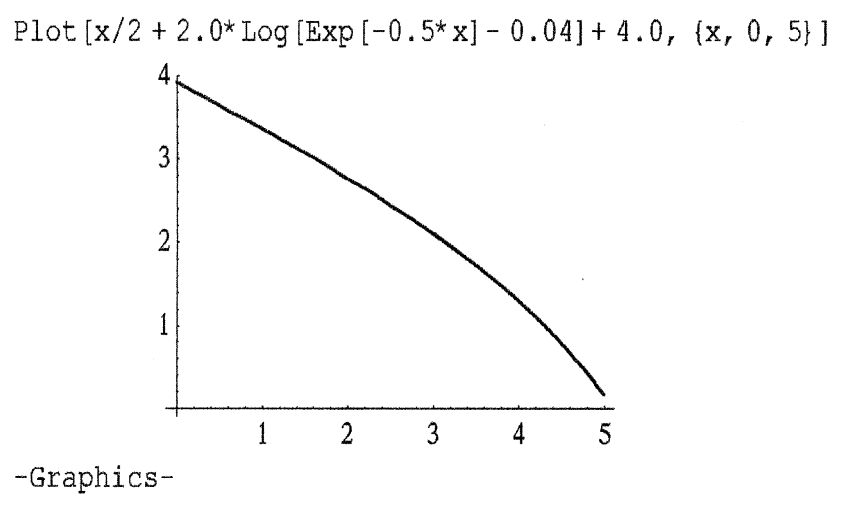

Fig.3

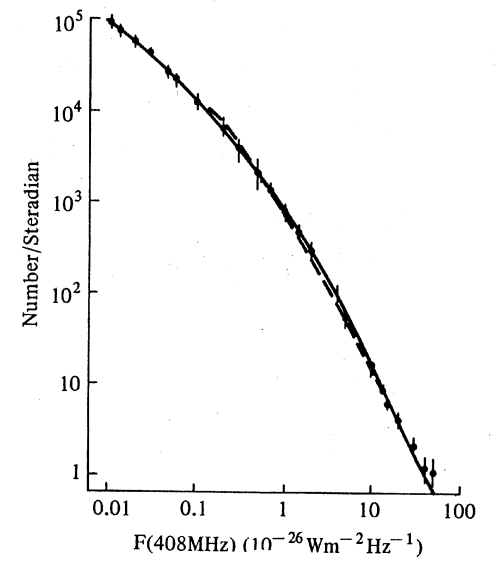

Fig. 4 (from [6])

curve, due to Ryle, is shown in Fig. 4 [7]. We cannot help admiring the excellent agreement of the two curves! Note that the actual curve of radio counts "is not consistent with static Euclidean space, the steady-state model, or simple Friedmann models" [6].

We remark that by appropriately fitting exactly the theoretical curve to the data by choosing the best values of $C_1$ & $C_2$ , we can find first $U_0$ from $C_1$. Then we can find G from $C_2$ as well. Also, as we will see in section 7, knowing $H_0$ (observable) we can determine the particle horizon, given by $U_0\sigma_0$ , by the relation

$$U_0\sigma_0 = c/H_0 . \tag{3.31}$$

Then, having found $U_0$ above, we can also determine $\sigma_0$, the "coordinate" particle horizon.

## 4. Hubble´s law

We work in the same approximation as in the preceding section.

The Hubble parameter is defined by

$$H = \frac{\dot{U}(t)}{U(t)}. \tag{4.1}$$

Substituting U(t) from (1.4), we obtain

$$H = G, \tag{4.2}$$

at all times.

The element of spatial distance dl is given by

$$dl = U(t)d\sigma . \tag{4.3}$$

The "instantaneous" finite distance at time t, will then be

$$l = U(t)\int d\sigma = U(t)\sigma . \tag{4.4}$$

For two galaxies separated by this distance, the rate of change of their distance, that is their relative "velocity" of recession, will be given by

$$v = \frac{dl}{dt} = \dot{U}(t)\sigma = \frac{\dot{U}}{U}(U\sigma) = \frac{\dot{U}}{U}l, \tag{4.5}$$

so that

$$v = Hl = Gl. \tag{4.6}$$

This is *Hubble´s law*. Note that this velocity is not a local concept, and so *it is not limited to be less than c.* It can be even superluminal!

But what we are actually observing is not the velocity v, but rather the redshift z of a galaxy. We can find the Hubble law in terms of z as follows.

We have

$$1+z=\frac{\lambda_0}{\lambda}, \tag{4.7}$$

where $\lambda_0$ is the wavelength of light coming from some (cosmological) source at the observer, and λ the wavelength of the same light at the source. But, because of the smallness of this wavelength, as compared with cosmological distances, we may set $\lambda_0 = dl_0$ & $\lambda = dl$. Then we have

$$\frac{\lambda_0}{\lambda}=\frac{U_0}{U}. \tag{4.8}$$

Writing $U = U(t_0 - (t_0 - t))$, with $t_0 - t$ being small, we have approximately, if we develop in Taylor series,

$$U\left(t_0-(t_0-t)\right)\cong U(t_0)-\frac{t_0-t}{1!}\dot{U}(t_0), \tag{4.9}$$

from which

$$1+z\cong\frac{1}{1-\frac{\dot{U}_0}{U_0}\Delta t}\cong 1+\frac{\dot{U}_0}{U_0}\Delta t, \tag{4.10}$$

and we obtain

$$cz\cong\frac{\dot{U}_0}{U_0}(c\Delta t)=H_0 d, \tag{4.11}$$

where $d = c\Delta t$ is the light distance.

We have met three kinds of distance: the instantaneous distance $l = U\sigma$ , the light distance $d = c\Delta t$, and the luminosity distance $D = \sqrt{(S/4\pi)}$ (*). But, since we work to the lowest order, it is evident that $l \cong d \cong D$. Also, in the same approximation, we observe that $cz \cong v$, resembling the Doppler shift. Of course it is *not* a Doppler shift, since the matter is *comoving* in our coordinate system.

Using the luminosity distance D, we can write for Hubble´s law

---

(*) The luminosity distance D is defined by the requirement that the flux density $F_0$ of a source with luminosity L is given by the relation $F_0 = L/(4\pi D)$.

$$cz \cong H_0 D = H_0 \sqrt{\frac{S}{4\pi}} = H_0 \sqrt{\frac{E_0}{4\pi E}}, \tag{4.12}$$

where $E_0$ is the luminosity of the source and E the flux density at the observer, so that

$$\log z \cong const. - ½ \log E \tag{4.13}$$

But, if we use instead of E the astronomical magnitude m, given by

$$m = const. - 2.5 \log E, \tag{4.14}$$

we find finally from (4.13)

$$\log z \cong const. - ½(-m/2.5), \tag{4.15}$$

or

$$\log z \cong const. + 0.2m \tag{4.16}$$

This is the relation giving the Hubble diagram.

## 5. The deceleration parameter and the distance-redshift relation

As we have seen, the Hubble parameter is given by

$$H = \dot{U}/U = G \geq 0 \tag{5.1}$$

The deceleration parameter is defined by

$$q = -\ddot{U}U/\dot{U}^2; \quad U(t) = \Theta E e^{Gt}. \tag{5.2}$$

We find

$$\dot{U}(t) = \Theta E G e^{Gt} = GU(t) \tag{5.3}$$

and

$$\ddot{U}(t) = G\dot{U}(t) = G^2 U(t). \tag{5.4}$$

Thus substituting U, $U^{\cdot}$, and $U^{\cdot\cdot}$ in (5.2), we find for the deceleration parameter

$$q = -\frac{G^2 U^2}{G^2 U^2} = -1. \tag{5.5}$$

This corresponds to the steady-state model, and it means that the Universe is actually *accelerating* its expansion.

The distance-redshift relation is now derived in a completely similar fashion with the Friedmann models [2]. It gives the redshift z versus the luminosity distance D, with parameters H & q, and it is given by

$$cz = HD + \frac{H^2}{2c}(q-1)D^2 + ... \tag{5.6}$$

This relation has to be satisfied approximately for q = -1, and H = G (we emphasize that we always take $\omega_1 \cong 0 \cong \omega_2$ up to now).

## 6. Anisotropic Universe

A very important issue concerning our model Universe is the issue of isotropy. Obviously it is *not* isotropic, not even homogeneous (the space-time). It can magnificently be manifested by the anisotropy of the Hubble flow, as we will see in this section. And perhaps it is the proper time for an introduction of an anisotropy in our real Universe, since lately there is plenty evidence of an anisotropy at large scales in the distribution of redshifts on the sky, and moreover of the temperature, the polarization, and the morphology of the 3K microwave radiation on the celestial sphere (see arXiv.org/abs/astro-ph/…). Of course this anisotropy is much smaller in magnitude, by about a factor of $10^{-3}$, than the one induced by the motion of the observer. It is however measurable, if we subtract the anisotropy resulted by our motion. We will see in the following theoretical treatment that the anisotropy we are speaking of shows up even in the first approximation, and is a result of the non-vanishing of $\omega_1$ & $\omega_2$.

Consider first that we observe a redshift in the direction of the unit vector **r**. We then have

$$1 + z = \lambda_0 / \lambda = dl_{r(0)} / dl_r = U_0 / U \cong 1 + G\tau, \tag{6.1}$$

where G stands for the Hubble constant, and $\tau$ is the travel time of light. We assume of course that the light comes tangent to the r-curve passing through the observer, with the result to take $d\varphi_1 = d\varphi_2$. Use is also made of formulae (A24) in the Appendix A. We see thus that in the case on hand we have

$$cz \cong GD, \tag{6.2}$$

that is exactly the Hubble law.

Consider next that we observe a redshift in the direction of the unit vector $\boldsymbol{\varphi}_1$. We then have

$$1+z = \lambda_0/\lambda = dl_{\varphi_1(0)}/dl_{\varphi_1} \cong \frac{U_0}{U}\sqrt{\frac{1+U_0^{\,2}\omega_{1(0)}^{\ \ 2}/c^2}{1+U^2\omega_1^{\ 2}/c^2}}, \tag{6.3}$$

if we use again formulae (A24) in the Appendix A. Developing to the first order in $\omega_1^2$ & $\omega_{1(0)}^2$ , we have

$$1+z \cong \frac{U_0}{U}\left(1+\frac{1}{2c^2}U_0^{\,2}\omega_{1(0)}^{\ \ 2}-\frac{1}{2c^2}U^2\omega_1^{\ 2}\right), \tag{6.4}$$

so that, if we expand in Taylor series to first order in $\tau$ around their present values both U and $\omega_1$ , we find

$$1+z \cong (1+G\tau)\left(1+\frac{1}{c^2}U_0^{\,2}\tau\omega_{1(0)}\dot{\omega}_{1(0)}+\frac{1}{c^2}\omega_{1(0)}^{\ \ 2}\tau U_0\dot{U}_0\right), \tag{6.5}$$

and performing the multiplication to the same (first) order in $\tau$, we get

$$1+z \cong 1+G\tau+\frac{1}{c^2}\tau\left(U_0^{\,2}\omega_{1(0)}\dot{\omega}_{1(0)}+\omega_{1(0)}^{\ \ 2}U_0\dot{U}_0\right), \tag{6.6}$$

from which

$$cz \cong GD+\frac{D}{c^2}U_0^2\left(\omega_{1(0)}\dot{\omega}_{1(0)}+G\omega_{1(0)}^{\ \ 2}\right), \tag{6.7}$$

or finally, because of (1.2),

$$cz \cong GD-2\frac{GD}{c^2}U_0^2\omega_{1(0)}^{\ \ 2}, \tag{6.7$'$}$$

where of course we can use again any kind of distance (instantaneous d., light d., or luminosity d.) in the place of D, since we work to the first order.

Consider finally the case along $\boldsymbol{\varphi}_2$ . We similarly find

$$cz \cong GD+\frac{D}{c^2}(U_0 r)^2\left(\omega_{2(0)}\dot{\omega}_{2(0)}+G\omega_{2(0)}^{\ \ 2}\right), \tag{6.8}$$

or finally, because of (1.3),

$$cz \cong GD-2\frac{GD}{c^2}U_0^2\omega_{2(0)}^{\ \ 2}r^2, \tag{6.8$'$}$$

where r is the coordinate distance of the observer from the center of universal rotation.

Obviously

$$(cz)_r > (cz)_{\varphi_1}, (cz)_{\varphi_2}, \tag{6.9}$$

an effect giving rise to anisotropy.

Finally, let us see how we can extract the large scale temperature anisotropy $\delta T_0/T_0$ of the CMB. If $\delta z$ is the deviation of the redshift from the exact Hubble law, then

$$\delta z = \delta(1+z) = \delta(U_0/U) = \delta(T/T_0) \cong -\frac{T}{T_0}\frac{\delta T_0}{T_0}, \tag{6.10}$$

or

$$\delta z \cong -(1+z)\frac{\delta T_0}{T_0}, \tag{6.11}$$

so that

$$\frac{\delta T_0}{T_0} \cong -\frac{\delta z}{1+z}. \tag{6.12}$$

We have to remark that perhaps the above anisotropy caused the fact of disagreement among almost all determinations of the Hubble constant even up to our days, with the result for the astronomer to have no absolutely reliable value of that constant at his disposal up to now.

## 7. A new time coordinate, the Hubble law, and the age of cosmos

If in the metric (1.6) we set for the scale factor

$$\Theta E e^{Gt} = cT/\sigma_0, \tag{7.1}$$

where $\sigma_0$ is the coordinate particle horizon today, we can introduce a new time coordinate T in the metric. From (7.1) we find

$$dt = \frac{1}{G}\frac{dT}{T}, \tag{7.2}$$

so that

$$c^2 dt^2 = c^2 \frac{1}{G^2}\frac{dT^2}{T^2}, \tag{7.3}$$

and the metric becomes

$$ds^2 = \frac{c^2}{G^2}\frac{dT^2}{T^2} - \frac{c^2T^2}{\sigma_0^2}\left(dr^2 + r^2 d\Phi^2 + d\phi^2\right). \tag{7.4}$$

An interresting feature of the new metric is the following. From (7.4) we get

$$g_{00} = \frac{1}{G^2T^2}, \tag{7.5}$$

so that

$$\sqrt{g_{00}} = \frac{1}{GT}, \tag{7.6}$$

and we find for the proper time

$$T > 0: \quad d\tau_+ = \sqrt{g_{00}}dT = \frac{dT}{GT} > 0 \tag{7.7}$$

$$T < 0: \quad d\tau_- = \sqrt{g_{00}}dT = \frac{dT}{GT} < 0 \tag{7.8}$$

Thus, for T > 0 we take our familiar Universe, which we call *cosmos,* while for T < 0 we take the *anti-cosmos* (cf. e-Appendix B in [1]). In other words, the Universe for T < 0 is the right place for anti-cosmos!

Concerning the *age* of cosmos, it is easy to see that it is infinite in terms of the coordinate t, or even of the proper time τ. However it is interesting to see that it is not the case for the new coordinate T (this was the underlying motivation for introducing T at the first place).

I f we denote by G the Hubble constant in the t-time from now on, and by H the Hubble parameter in the T-time, we have (denoting by a prime (′) the derivation with respect to T)

$$H \equiv \frac{U'}{U} = G\frac{dt}{dT}, \tag{7.9}$$

or, because of (7.2),

$$H = G\frac{1}{GT} = \frac{1}{T}. \tag{7.10}$$

The same result, because of (7.1), is achieved in the fashion

$$H \equiv \frac{U'}{U} = \frac{\left(\frac{cT}{\sigma_0}\right)'}{\left(\frac{cT}{\sigma_0}\right)} = \frac{1}{T}. \tag{7.11}$$

We can easily see , either from (7.10), or from (7.11), that the *age* of Cosmos in the T-time is

$$T_0 = 1/H_0 . \tag{7.12}$$

Concerning the Hubble law, it continues to hold in the new time, since

$$D = U\sigma , \tag{7.13}$$

as

$$V \equiv \frac{dD}{dT} = U'\sigma = \frac{U'}{U}(U\sigma) = HD. \tag{7.14}$$

And because we actually observe the redshift z rather than the velocity V, we have

$$1 + z = \lambda_0/\lambda = dl_0/dl . \tag{7.15}$$

But

$$dl = \frac{cT}{\sigma_0} d\sigma \tag{7.16}$$

and

$$dl_0 = \frac{cT_0}{\sigma_0} d\sigma . \tag{7.17}$$

Thus (7.15) gives

$$1 + z = T_0/T , \tag{7.18}$$

or

$$z = \Delta\lambda/\lambda = \Delta T/T . \tag{7.19}$$

From (7.10), or (7.11), we then find

$$cz = H(c\Delta T) = HD, \tag{7.20}$$

which has to be compared with (7.14). Alternatively, from ds = 0 for light, we get

$$\frac{cdT}{GT} = \frac{cT}{\sigma_0} d\sigma \tag{7.21}$$

But

$$d\tau = \sqrt{g_{00}} dT = \frac{dT}{GT}. \tag{7.22}$$

Thus

$$cd\tau = \frac{cT}{\sigma_0} d\sigma = dl, \qquad (7.23)$$

which is OK. So we have again, from (7.16) & (7.17)

$$1 + z = \nu / \nu_0 = d\tau_0 / d\tau = T_0 / T \qquad (7.24)$$

(cf. with (7.18)). Note that from

$$\lambda \nu = c \qquad (7.25)$$

we get differentiating

$$d\lambda / \lambda = -d\nu / \nu . \qquad (7.26)$$

Thus

$$z = -\frac{d\nu}{\nu} \cong \frac{\nu}{\nu_0} - 1, \qquad (7.27)$$

which we insert into (7.24).

Finally, we have to justify why $T_0$ gives in fact the *age* of the Universe. This can be easily done by noting that the particle horizon, which is given by the proper distance $U_0\sigma_0$ , has to be set equal to the actual age of the Universe times the velocity of light c. We have in fact then

$$(age) = \frac{1}{c} U_0 \sigma_0 = \frac{\sigma_0}{c} \frac{cT_0}{\sigma_0} = T_0, \qquad (7.28)$$

And thus we correctly take this age from formulae (7.10), and/or (7.11), as given by

$$T_0 = 1/H_0 . \qquad (7.29)$$

Note that $H_0$ is observable. Thus we can find $T_0$ from $H_0$ .

**Note concerning non-cosmological implications**

It has to be noted that of course non-cosmological implications of the present model Universe exist as well. We can distinguish among them the implications on the structure and dynamical evolution of galaxies. We mention in particular two of them.

First, the explanation of the galactic flat rotation curves, attributed to the differential rotation of the Universe. See physics/0602035.

Second, the explanation of the spiral structure in spiral galaxies, attributed to Coriolis forces, coming from the rigid rotation of the Universe. See physics/0607214.

In these two applications we have assumed the constant metric (G = 0), but the analysis remains valid *approximately* even when we assume the general metric(G ≠ 0) of the present paper. It is sufficient to assume in addition that the field changes *slowly,* or that the analysis applies for a *short* period of time.

**Appendix A: The spatial metric**

We will use the notation of [4].

We define the spatial metric (three-)tensor $\gamma_{\alpha\beta}$ in such a way that the elementary proper length dl is given by

$$dl^2 = \gamma_{\alpha\beta} dx^\alpha dx^\beta \qquad (\alpha, \beta = 1, 2, 3). \tag{A1}$$

We have

$$\gamma^{\alpha\beta} = -g^{\alpha\beta}, \tag{A2}$$

and

$$\gamma_{\alpha\beta} = -g_{\alpha\beta} + \frac{g_{0\alpha} g_{0\beta}}{g_{00}}. \tag{A3}$$

Of course

$$g_{\alpha\beta} = \begin{pmatrix} -U^2 & & 0 \\ & -U^2 & \\ 0 & & -U^2 r^2 \end{pmatrix}. \tag{A4}$$

We also define

$$g_\alpha = -\frac{g_{0\alpha}}{g_{00}}, \tag{A5}$$

hence

$$g^\alpha = -g^{0\alpha}. \tag{A6}$$

For the metric (4-)tensor we have for the case on hand

$$g_{00} = 1 - U^2 \frac{\omega_1^2}{c^2} - U^2 r^2 \frac{\omega_2^2}{c^2} \quad \text{(A7.1)}$$

$$g_{01} = 0 \quad \text{(A7.2)}$$

$$g_{02} = -U^2 \frac{\omega_1}{c} \quad \text{(A7.3)}$$

$$g_{03} = -U^2 r^2 \frac{\omega_2}{c} \quad \text{(A7.4)}$$

Thus, denoting $g_{00}$ by h, we have from (A5)

$$g_\alpha = \frac{1}{h}\left(0, U^2 \frac{\omega_1}{c}, U^2 r^2 \frac{\omega_2}{c}\right). \quad \text{(A8)}$$

For the covariant metric tensor we have

$$g_{ik} = \begin{pmatrix} g_{00} \cdots g_{0\alpha} \cdots \\ \vdots \\ g_{\alpha 0} \quad g_{\alpha\beta} \\ \vdots \end{pmatrix} = \begin{pmatrix} 1 - U^2 \frac{\omega_1^2}{c^2} - U^2 r^2 \frac{\omega_2^2}{c^2}, 0, -U^2 \frac{\omega_1}{c}, -U^2 r^2 \frac{\omega_2}{c} \\ 0 & -U^2 & 0 & 0 \\ -U^2 \frac{\omega_1}{c} & 0 & -U^2 & 0 \\ -U^2 r^2 \frac{\omega_2}{c} & 0 & 0 & -U^2 r^2 \end{pmatrix} \quad \text{(A9)}$$

We will first find the reciprocal tensor $g^{ik}$ , given by

$$g^{ik} = \left(\frac{G_{ki}}{g}\right), \quad \text{(A10)}$$

where g is the determinant of $g_{ik}$ and $G_{ki}$ the coefficient of $g_{ik}$ in the development of g.

Thus, we need g and then $G_{ki}$ . We find

$$g = -U^6 r^2. \quad \text{(A11)}$$

Also then

$$G_{00} = -U^6 r^2 \quad \text{(A12.1)}$$

$$G_{01} = G_{10} = 0 \quad \text{(A12.2)}$$

$$G_{02} = G_{20} = U^6 r^2 \frac{\omega_1}{c} \quad \text{(A12.3)}$$

$$G_{03}=G_{30}=U^6 r^2\frac{\omega_2}{c} \tag{A12.4}$$

Thus

$$g^{00}=1 \tag{A13.1}$$

$$g^{01}=0 \tag{A13.2}$$

$$g^{02}=-\frac{\omega_1}{c} \tag{A13.3}$$

$$g^{03}=-\frac{\omega_2}{c}. \tag{A13.4}$$

We also find

$$G_{11}=U^4 r^2 \tag{A14.1}$$

$$G_{22}=U^4 r^2\left(1-U^2\frac{\omega_1^{\;2}}{c^2}\right) \tag{A14.2}$$

$$G_{33}=U^4\left(1-U^2 r^2\frac{\omega_2^{\;2}}{c^2}\right) \tag{A14.3}$$

$$G_{12}=G_{13}=0 \quad ; \quad G_{23}=U^6 r^2\frac{\omega_1\omega_2}{c^2} \tag{A14.4}$$

Thus, we obtain

$$g^{11}=-\frac{1}{U^2} \tag{A15.1}$$

$$g^{22}=-\frac{1-U^2\frac{\omega_1^{\;2}}{c^2}}{U^2} \tag{A15.2}$$

$$g^{33}=-\frac{1-U^2 r^2\frac{\omega_2^{\;2}}{c^2}}{U^2 r^2} \tag{A15.3}$$

$$g^{12}=g^{13}=0 \quad ; \quad g^{23}=-\frac{\omega_1\omega_2}{c^2} \tag{A15.4}$$

Then, for $\gamma_{\alpha\beta}$, we find from (A3)

$$\gamma_{11} = U^2 \tag{A16.1}$$

$$\gamma_{22} = U^2 + \frac{U^4 \frac{\omega_1^2}{c^2}}{1 - U^2 \frac{\omega_1^2}{c^2} - U^2 r^2 \frac{\omega_2^2}{c^2}} \tag{A16.2}$$

$$\gamma_{33} = U^2 r^2 + \frac{U^4 r^4 \frac{\omega_2^2}{c^2}}{1 - U^2 \frac{\omega_1^2}{c^2} - U^2 r^2 \frac{\omega_2^2}{c^2}} \tag{A16.3}$$

$$\gamma_{12} = \gamma_{13} = 0 \tag{A16.4}$$

$$\gamma_{23} = \frac{U^4 r^2 \frac{\omega_1 \omega_2}{c^2}}{1 - U^2 \frac{\omega_1^2}{c^2} - U^2 r^2 \frac{\omega_2^2}{c^2}}. \tag{A16.5}$$

Also, $g_\alpha$ as given by (A5) yields

$$g_1 = 0 \tag{A17.1}$$

$$g_2 = \frac{U^2 \frac{\omega_1}{c}}{1 - U^2 \frac{\omega_1^2}{c^2} - U^2 r^2 \frac{\omega_2^2}{c^2}} \tag{17.2}$$

$$g_3 = \frac{U^2 r^2 \frac{\omega_2}{c}}{1 - U^2 \frac{\omega_1^2}{c^2} - U^2 r^2 \frac{\omega_2^2}{c^2}}. \tag{A17.3}$$

For the contravariant components, given by (A.6), we find

$$g^1 = 0 \tag{A18.1}$$

$$g^2 = \frac{\omega_1}{c} \tag{A18.2}$$

For the contravariant spatial metric $\gamma^{\alpha\beta}$, given by (A2), we find

$$\gamma^{11} = \frac{1}{U^2} \tag{A19.1}$$

$$\gamma^{22} = \frac{1 - U^2 \frac{\omega_1^2}{c^2}}{U^2} \tag{A19.2}$$

$$\gamma^{33} = \frac{1 - U^2 r^2 \frac{\omega_2^2}{c^2}}{U^2 r^2} \tag{A19.3}$$

$$\gamma^{12} = \gamma^{13} = 0 \quad ; \quad \gamma^{23} = \frac{\omega_1 \omega_2}{c^2} \tag{A19.4}$$

For the determinant $\gamma$ we have

$$\gamma = -g/g_{00}\,, \tag{A20}$$

so that we find

$$\gamma = \frac{U^6 r^2}{1 - U^2 \frac{\omega_1^2}{c^2} - U^2 r^2 \frac{\omega_2^2}{c^2}}\,. \tag{A21}$$

For the proper volume, given by

$$dV = \sqrt{\gamma}\, dx^1 dx^2 dx^3, \tag{A22}$$

we get thus

$$dV = \frac{U^3 r}{\sqrt{1 - U^2 \frac{\omega_1^2}{c^2} - U^2 r^2 \frac{\omega_2^2}{c^2}}}\, dr\, d\varphi_1 d\varphi_2\,. \tag{A23}$$

We will need approximations for $\gamma_{11}$, $\gamma_{22}$, $\gamma_{33}$. To first order in $\omega_1^2$ & $\omega_2^2$, we find

$$\gamma_{11} = U^2 \tag{A24.1}$$

$$\gamma_{22} \cong U^2 \left( 1 + U^2 \frac{\omega_1^2}{c^2} \right) \tag{A24.2}$$

$$\gamma_{33} \cong U^2 r^2 \left( 1 + U^2 r^2 \frac{\omega_2^2}{c^2} \right) \tag{A24.3}$$

**Appendix B: The shape and the equations of the Universe**

**1.** One may ask: how can we identify the (r, Φ) planes for z = 0 & z = 2π (see fig. 2), since all (r, Φ) planes remain parallel to one another. The answer is that we can do it if we imagine our three-dimensional space embedded in a fictitious four-dimensional (Euclidean) space. (Thus we return to the picture from where we started in the beginning). To make things clear, suppose that we suppress the Z coordinate on these planes, and we retain only one dimension, say x (x ⊥ Z), from each of these planes. Then our three-dimensional space will become a two-dimensional space, (x, z), which we can imagine embedded in a three-dimensional *Euclidean* space (not necessarily in the former one). As long as we move on the two dimensional space (x, z), we may consider it flat (its metric is indeed flat). If however take in mind the third dimension of the Euclidean space mentioned, we can see that in reality the line segment [0, 2π) must be curved in the circumference of a circle, with center *outside* the space (x, z) and unit radius, perpendicular to all x´s axes (being parallel to one another) in a way which permits us to identify 0 with 2π. The two-dimensional space will thus be in reality the surface of an infinite circular cylinder. Its metric permits it, since there is no difference in the line element on a plane and on a cylinder. Now suppose that instead of the straight lines x we have the planes (r, Φ) ≡ (x, Z), and instead of the two-dimensional space (x,z) we have our three-dimensional space. Then we will need a fictitious *Euclidean* four-dimensional space instead of the above three-dimensional *Euclidean* space, in order to perform the same process.

It is now evident that the origin O, which has been taken as the center of differential rotation on the φ´s plane (see fig. 1) must be taken *outside* our three-dimensional space (C). In this respect the earlier picture of the center of differential rotation belonging in our three-dimensional space is naïve. It is simply as if we attributed a rotational motion of the x-axes around O *on* the surface (x, z), which we imagine as a plane and not as a cylindrical surface. Only the Φ-rotation on an (r, Φ) plane has its center, say O, in our three-dimensional space (see fig. 1). Indeed, in this case the Φ-rotations of all (r, Φ) planes will take place around those centers, which form the Oz axis (see fig. 2) from 0 to 2π. Of course the line segment [0, 2π) will be the circumference of a circle of unit radius, as explained above.

Thus, from the above considerations we see that: 1) the surfaces of constant φ are the (r, Φ) planes, 2) the surfaces of constant Φ are truncated conical, because of the φ rotation, with a common axis Cx, passing the latter through the center C of Universal rotation (lying *outside* our Universe) (see fig.5), and parallel to the direction Ox, & 3) the surfaces of constant r will be tori with the common center of the torus′ great circle identified with the center of Universal (φ-)rotation (C), and the centers of the small circles along the line segment [0, 2π) curved to the circumference O (see fig. 6). As a conclusion we may say with confidence that *the shape of our Universe is hypertoroidal.*

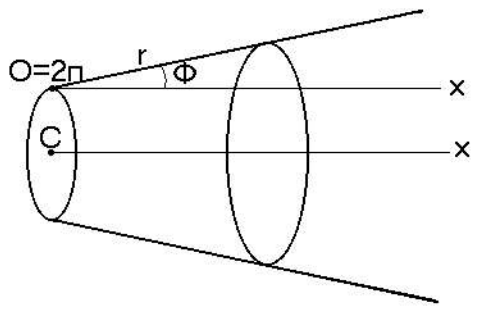

Fig. 5

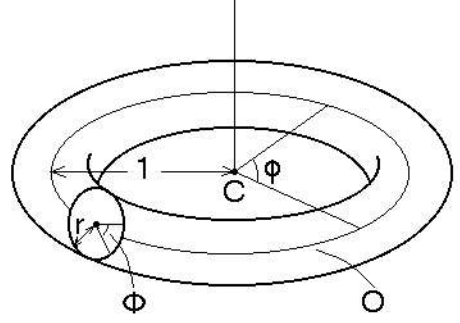

Fig. 6

This is the case mainly because the orbit of a particle of the cosmic fluid, considered earlier as a solid helix, is a helix on the surface of a torus. Thus, a particle of the cosmic fluid will remain on the surface of one and the same member of a family of tori, because right r is a commoving coordinate and does not change for the particle on hand. And on this toroidal surface it will perform the two rotational motions: one around the O curve (circle) (see fig. 6) with angular velocity $\omega_2 = d\Phi/dt$, and the other around the center C of Universal rotation (lying outside our three-dimensional space) with angular velocity $\omega_1 = d\varphi/dt$.

We can take the torus mentioned above from topological considerations, too. We know that a torus can be taken from a square by identifying its opposite sides, one pair after the other. After the first identification we will take a cylindrical surface limited by two end-circles, which are the other pair of sides of the square curved until they close to circles. Thus, the second identification concerns these two limiting circles, and it leads right to the torus.

Now, in our case the cylindrical surface already exists. It is the cylindrical surface with axis Oz and radius r, limited by its circles at $z = 0$ and $z = 2\pi$ (see fig. 2). It is then simple to identify these circles and take topologically the torus.

Then, of course the various cylinders for different r´s, as being “one inside the other”, will result in a family of tori “one inside the other”, our Universe.

So far we have examined the purely spatial part $d\sigma^2$ of the metric. If we take in mind also the scale factor U(t), we will see that our 3-dimensional space will expand accelerating, with the angular velocities $\omega_1$ & $\omega_2$ relaxing. As we go back in time following the shrinking of the Universe, we will approach at the limit the center C, where the big-bang took place.

The total volume of the hypertorus (of our Universe) will be obviously infinite concerning the metric $d\sigma^2$, since it contains points lying at infinity ($r = \infty$). But concerning the spatial metric $dl^2 = U^2(t)d\sigma^2$ at some instance $t_0$ the volume of the Universe will be finite. In fact, we have to take (for negligible $\omega_1$ & $\omega_2$)

$$V = (2\pi)^2 \int_0^R U^3(t) r dr \tag{B1}$$

or

$$V = (2\pi)^2 \frac{c^2}{2G^2} U_0 \frac{R^2}{(R+\alpha)^2} \tag{B2}$$

(cf. (3.6) & (3.12)) for $R \to \infty$, so that

$$V = 2\pi^2 \frac{c^2}{G^2} U(t_0). \tag{B3}$$

**2.** Another approach, more formal, to find the shape of the Universe is to find directly the equations which describe it, as follows. If we imagine our Universe embedded in the fictitious Euclidean four-dimensional space ($x_1$, $x_2$, $x_3$, $x_4$) from which we started our investigation at the first place (see [1], pp. 11-12), the equation (35) of [1], which gives the line element of this four-space, if we take in mind the equations of p. 52 of [1], becomes (remember that we have renamed z to r)

$$dL^2 = \left(\frac{\partial a_1}{\partial C} dC\right)^2 + U^2(t) d\phi^2 + \left(U(t) dr + \frac{\partial a_2}{\partial C} dC\right)^2 + U^2(t) r^2 d\Phi^2, \tag{B4}$$

from which we take the line element $dl^2$ of the hypersurface, which is our Universe, by letting $dC = 0$, so that

$$dl^2 = U^2(t)\{dr^2 + r^2 d\Phi^2 + d\phi^2\}, \tag{B5}$$

in complete agreement with (1.6), which gives (1.9). Thus our results are first verified.

Next, if we take in mind eq. (305) of [1], namely

$$\left.\begin{aligned} a_1 &= U(t) \\ a_2 &= rU(t) \end{aligned}\right\}, \tag{B6}$$

the eqns. (34) of [1] giving the hypersurface (our Universe) in parametric form (disregarding the scale factor for the moment) become

$$\left.\begin{aligned} x_1 &= \cos\phi \\ x_2 &= \sin\phi \\ x_3 &= r\cos\Phi \\ x_4 &= r\sin\Phi \end{aligned}\right\} \tag{B7}$$

We see at once that, for the comoving coordinate r fixed (thus for two parameters: φ & Φ), eqns. (B7) describe a torus in four (4) dimensions, since they describe the Cartesian product of two circles.

If we write (B7) in the form

$$\left.\begin{aligned} x_1 &= \cos(\varphi_1 + \omega_1 t) \\ x_2 &= \sin(\varphi_1 + \omega_1 t) \\ x_3 &= r\cos(\varphi_2 + \omega_2 t) \\ x_4 &= r\sin(\varphi_2 + \omega_2 t) \end{aligned}\right\} \tag{B8}$$

we have the equations of the orbit of a particle of the cosmic fluid in parametric form. The section by the plane ($x_3$, $x_4$) gives an harmonic motion. The same applies to the ($x_1$, $x_2$) section, giving another harmonic motion. Finally the ($x_1$, $x_3$), ($x_1$, $x_4$), ($x_2$, $x_3$), ($x_2$, $x_4$) sections give simply Lissajous pictures. Of course, in the case of the orbits the scale factor U(t) has to be also taken into account.

The area of this torus is obviously (see (B5), where we disregard U(t), set dr = 0, and we apply the Pythagorean theorem)

$$S = \iint r d\Phi d\phi = (2\pi)^2 r. \tag{B9}$$

Now, allow r to come into play. We then will have a family of tori, parametrized by r, which have common great circle of radius 1 (unity) and “cross-section” a circle of radius r, in a way being “one inside another”. In this sense we may, thus, say that our Universe is a “hypertorus” with the meaning described above as a family of tori.

If finally we take also in mind the scale factor U(t), the resulting situation is easy to be understood, and has been described in **1.** above.

**Conclusion**

As a conclusion, we may say about the shape of the Universe that, for a given instant of time $t_0$ , it is a hypertorus, that is a family of toroidal surfaces, "built" upon the circumference of a circle with radius $U(t_0)$. All this has to be meant as accelerating expanding with time t, in such a way that the family remains similar to itself. We have even found the equations giving the shape of the hypersurface representing the Universe, which are right eqns. (B7).